\documentclass[english]{article}
\usepackage[T1]{fontenc}
\usepackage[latin9]{inputenc}

\makeatletter

\newcommand{\lyxaddress}[1]{
\par {\raggedright #1
\vspace{1.4em}
\noindent\par}
}

\usepackage{babel}
\makeatother

\begin{document}

\title{Nearly Invariance of Jarlskog Determinant due to Planck Scale Effects }

\author{Bipin Singh Koranga}

\maketitle

\lyxaddress{Kirori Mal college (University of Delhi,) Delhi-110007, India}

\begin{abstract}
An asymmetry between the probabilities $P(\nu_{\mu}\rightarrow\nu_{e})$
and $P(\bar{\nu_{\mu}}\rightarrow\bar{\nu_{e}})$ would be direct
indication of CP violation at the fundamental level. Planck scale
effects on neutrino mixing, we have derived the mixing angles of neutrino
flavour due to Planck scale effects. It has been shown that Jarlskog
determinant remains nearly invariant due to Planck scale effects.
\end{abstract}

\section{INTRODUCTION}

The evidence of a deficit of detected solar neutrinos {[}1] indicates
that electron neutrinos must also participate in lepton mixing. The
participation of all three neutrinos in lepton mixing raises the possibility
of CP and T violation in neutrino oscillations. The emergence of large
mixing parameter in lepton sector indicate the potentially large CP
and T violation are maximal for neutrinos in vacuum. A number of auther
{[}2-6] have explored the phenomenology of CP and T violation in neutrino
oscillations for several different scenarios of lepton masses and
mixing parameters. In both solar and atmospheric neutrino experiments
can traverse a significant of the earth. Long baseline accelerator
and reactor experiment still longer baseline. CP violation arise as
three or more generation {[}7, 8]. CP violation in neutrino oscillation
is interesting because it relates directly to CP phase parameter in
the mixing for $n>3$ degenerate neutrino. We can write down the compact
formula for the difference of transition probability between conjugate
channel.

\begin{equation}
\Delta P(\alpha,\beta)=P(\nu_{\mu}\rightarrow\nu_{e})-P(\bar{\nu_{\mu}}\rightarrow\bar{\nu_{e}}),\end{equation}

where \[
(\alpha,\beta)=(e,\mu),(\mu,\tau),(\tau,e).\]

The main physical goal in future experiment are the determination
of the unknown parameter $\theta_{13}$ and upper bound $sin^{2}2\theta_{13}<0.01$
is obtained for the ref {[}9]. In particular, the observation of $\delta$
is quites interesting for the point of view that~$\delta$ related
to the origin of the matter in the universe. The determination of
$\delta$ is the final goal of the future experiments. We get the
analytical expression for $\Delta P(\alpha,\beta)$ using the usual
form of the MNS matrix parametrization {[}10].

\begin{equation}
U=\left(\begin{array}{ccc}
c_{12}c_{13} & s_{12}c_{13} & s_{13}e^{-i\delta}\\
-s_{12}c_{23}-c_{12}s_{23}s_{13}e^{i\delta} & c_{12}c_{23}-s_{12}s_{23}s_{13}e^{i\delta} & s_{23}c_{13}\\
s_{12}s_{23}-c_{12}c_{23}s_{13}e^{i\delta} & -c_{12}s_{23-}s_{12}s_{13}s_{23}e^{i\delta} & c_{23}s_{13}\end{array}\right),\end{equation}

where c and s denoted the cosine and sine of the respective notation,
thus $\Delta P(\alpha,\beta)$ in vacuum can be written as\begin{equation}
\Delta P(\alpha,\beta)=16J\left(sin\Delta_{21}sin\Delta_{32}sin\Delta_{31}\right).\end{equation}

Here $\alpha$ and $\beta$ denote different neutrino or anti-neutrino
flavour

where

\begin{equation}
\Delta_{ij}=1.27\left(\frac{\Delta_{ij}}{eV^{2}}\right)\left(\frac{L}{Km}\right)\left(\frac{1GeV}{E}\right),\end{equation}

$\Delta_{ij}=(m_{i}^{2}-m_{j}^{2})$is the difference of $i^{th}$
and $j^{th}$ vacuum mass square eigen value, E is the neutrino energy
and L is the travel distance and the well known Jarlskog determinant
{[}11], J is the standard mixing parametrization is given by

\[
J=Im\left(U_{e1}U_{e2}^{*}U_{\mu1}^{*}U_{\mu2}\right)\]

\begin{equation}
=\frac{1}{8}sin2\theta_{12}sin2\theta_{23}sin2\theta_{13}cos\theta_{13}sin\delta,\end{equation}

and the asymmetry parameter suggested by Cabibbo {[}12], as an alternative
to measure CP violation in the lepton sector

\begin{equation}
A_{cp}=\frac{\Delta P}{P(\nu_{\mu}\rightarrow\nu_{e})-P(\bar{\nu_{\mu}}\rightarrow\bar{\nu_{e}}),}\end{equation}

The purpose of this paper is to study the Planck scale effects on
Jarlskog Determinant. In Sec-2, we discuss the neutrino mixing angle
due to Planck scale effects. In Sec-3, we give the conclusions.

\section{NEUTRINO MIXING ANGLE DUE TO PLANCK SCALE EFFECTS}

To calculate the effects of perturbation on neutrino observables.
The calculation developed in an earlier paper {[}13]. A natural assumption
is that unperturbed ($0^{th}$ order mass matrix $M$~is given by

\begin{equation}
\mathbf{M}=U^{*}diag(M_{i})U^{\dagger},\end{equation}

where, $U_{\alpha i}$ is the usual mixing matrix and $M_{i}$ , the
neutrino masses is generated by Grand unified theory. Most of the
parameter related to neutrino oscillation are known, the major expectation
is given by the mixing elements $U_{e3}.$ We adopt the usual parametrization.

\begin{equation}
\frac{|U_{e2}|}{|U_{e1}|}=tan\theta_{12}\end{equation}

\begin{equation}
\frac{|U_{\mu3}|}{|U_{\tau3}|}=tan\theta_{23}\end{equation}

\begin{equation}
|U_{e3}|=sin\theta_{13}\end{equation}

In term of the above mixing angles, the mixing matrix is

\begin{equation}
U=diag(e^{if1},e^{if2},e^{if3})R(\theta_{23})\Delta R(\theta_{13})\Delta^{*}R(\theta_{12})diag(e^{ia1},e^{ia2},1).\end{equation}

The matrix $\Delta=diag(e^{\frac{1\delta}{2}},1,e^{\frac{-i\delta}{2}}$)
contains the Dirac phase. This leads to CP violation in neutrino oscillation
$a1$ and $a2$ are the so called Majoring phase, which effects the
neutrinoless double beta decay. $f1,$ $f2$ and $f3$ are usually
absorbed as a part of the definition of the charge lepton field. Planck
scale effects will add other contribution to the mass matrix that
gives the new mixing matrix can be written as {[}13]

\[
U^{'}=U(1+i\delta\theta),\]

\begin{equation}
=\left(\begin{array}{ccc}
U_{e1} & U_{e2} & U_{e3}\\
U_{\mu1} & U_{\mu2} & U_{\mu3}\\
U_{\tau1} & U_{\tau2} & U_{\tau3}\end{array}\right)+i\left(\begin{array}{ccc}
U_{e2}\delta\theta_{12}^{*}+U_{e3}\delta\theta_{23,}^{*} & U_{e1}\delta\theta_{12}+U_{e3}\delta\theta_{23}^{*}, & U_{e1}\delta\theta_{13}+U_{e3}\delta\theta_{23}^{*}\\
U_{\mu2}\delta\theta_{12}^{*}+U_{\mu3}\delta\theta_{23,}^{*} & U_{\mu1}\delta\theta_{12}+U_{\mu3}\delta\theta_{23}^{*}, & U_{\mu1}\delta\theta_{13}+U_{\mu3}\delta\theta_{23}^{*}\\
U_{\tau2}\delta\theta_{12}^{*}+U_{\tau3}\delta\theta_{23}^{*}, & U_{\tau1}\delta\theta_{12}+U_{\tau3}\delta\theta_{23}^{*}, & U_{\tau1}\delta\theta_{13}+U_{\tau3}\delta\theta_{23}^{*}\end{array}\right).\end{equation}

Where $\delta\theta$ is a hermition matrix that is first order in
$\mu${[}13,14]. The first order mass square difference $\Delta M_{ij}^{2}=M_{i}^{2}-M_{j}^{2},$get
modified {[}13,14] as

\begin{equation}
\Delta M_{ij}^{'^{2}}=\Delta M_{ij}^{2}+2(M_{i}Re(m_{ii})-M_{j}Re(m_{jj})).\end{equation}

The change in the elements of the mixing matrix, which we parametrized
by $\delta\theta${[}13], is given by

\begin{equation}
\delta\theta_{ij}=\frac{iRe(m_{jj})(M_{i}+M_{j})-Im(m_{jj})(M_{i}-M_{j})}{\Delta M_{ij}^{'^{2}}}.\end{equation}

The above equation determine only the off diagonal elements of matrix
$\delta\theta_{ij}$. The diagonal element of $\delta\theta_{ij}$
can be set to zero by phase invariance. Using Eq(12), we can calculate
neutrino mixing angle due to Planck scale effects,

\begin{equation}
\frac{|U_{e2}^{'}|}{|U_{e1}^{'}|}=tan\theta_{12}^{'}\end{equation}

\begin{equation}
\frac{|U_{\mu3}^{'}|}{|U_{\tau3}^{'}|}=tan\theta_{23}^{'}\end{equation}

\begin{equation}
|U_{e3}^{'}|=sin\theta_{13}^{'}\end{equation}

As one can see from the above expression of mixing angle due to Planck
scale effects, depends on new contribution of mixing $U^{'}=U(1+i\delta\theta).$
To see the mixing angle due to Planck scale effects {[}13,15] only
$\theta_{13}$and $\theta_{12}$ mixing angle have small deviation
due to Planck scale effects.

\section{JARLSKOG DETERMINANT DUE TO PLANCK SCALE EFFECTS }

Let us compute Jarlskog determinant due to new mixing due to Planck
scale effects

\[
J^{'}=Im\left(U_{e1}^{'}U_{e2}^{'*}U_{\mu1}^{'*}U_{\mu2}^{'}\right)\]

\[
=Im((U_{e1}+i(U_{e2}\delta\theta_{12}^{*}+U_{e3}\delta\theta_{13}))((U_{e2}-i(U_{e1}^{*}\delta\theta_{12}^{*}+U_{e3}^{*}\delta\theta_{13}))\]

\begin{equation}
((U_{\mu1}^{*}-i(U_{\mu2}\delta\theta_{12}+U_{\mu3}\delta\theta_{13}))((U_{\mu2}+i(U_{\mu1}\delta\theta_{12}+U_{\mu3}\delta\theta_{23}^{*})\end{equation}

We simplified Jarlskog determinant due to new mixing matrix

\[
J^{'}=Im\left(U_{e1}U_{e2}^{*}U_{\mu1}^{*}U_{\mu2}\right)+Im(i(U_{\mu1}U_{\mu2})(|U_{e2}|^{2}\delta\theta_{12}^{*}+U_{e2}U_{e3}\delta\theta_{13}-|U_{e1}|^{2}\delta\theta_{12}^{*}-U_{e1}U_{e3}^{*}\delta\theta_{23}^{*})\]

\[
+Im(i(U_{e1}^{*}U_{e2})(|U_{\mu1}|^{2}\delta\theta_{12}+U_{\mu1}^{*}U_{\mu3}\delta\theta_{23}^{*}-|U_{\mu2}|^{2}\delta\theta_{12}-U_{\mu2}U_{\mu3}^{*}\delta\theta_{13}^{})\]

\[
=J+\Delta J\]

In term of mixing angle, we can write Jarlskog determinant in term
of mixing parameter due to Planck scale effects

\begin{equation}
J^{'}=\frac{1}{8}sin2(\theta_{12}+\epsilon_{12})sin2(\theta_{23}+\epsilon_{23})sin2(\theta_{13}+\epsilon_{13})cos(\theta_{13}+\epsilon_{13})sin\delta,\end{equation}

Numerically \% change of Jarlskog determinant is very small {[}17],
due to small change of mixing angle $\theta_{12}$ and $\theta_{13}.$
In reminder of this paper, we explore the neutrino mixing and jarlskog
determinant due to Planck scale effects. In order to ensure that our
discussion is relevant to theoretical, we will restrict our considerations
to neutrino mixing parameter due to Planck scale effects, which are
not experimental. Finally, we can wish one comment, if future experiment
find the non zero value of CP phase so we can say there is small possible
CP asymmetry due to Planck scale effects.


\begin{thebibliography}{10}
\bibitem{key-1}Supper-Kamiokande Collaboration, Y. Fukuda $et\, al.,$Phys.Lett
B \textbf{436}, 33, (1998).

\bibitem{key-1}G. C. Branco, M.N Rebelo, Acta Phys.Polon.\textbf{B}38:3819-3850,2007.

\bibitem{key-3}Hiroshi Nunokawa, Stephen J. Parke, Jose W.F. Valle,
Prog.Part.Nucl.Phys.\textbf{60}:338-402,2008.

\bibitem{key-4}Frans R. Klinkhamer, Phys.Rev.\textbf{D}73:057301,2006. 

\bibitem{key-5}M.C. Gonzalez $et\, al.,$Phys.Rev.\textbf{D}64:096006,2001

\bibitem{key-6}Zhi-zhong Xing, Phys.Lett.\textbf{B}487:327-333,2000.

\bibitem{key-7}M. Kobayashi and T. Maskawa, Prog. Theor. Phys. Rev.
Lett. \textbf{45}, 652 (1973).

\bibitem{key-8}V. Varger $et\, al.,$Phys.Rev.Lett \textbf{45}, 2084
(1980).

\bibitem{key-9}CHOOZ Collaboration, M. Apollonio, Phy. Lett B \textbf{420},
397 (1998).

\bibitem{key-10}Review of Particle Physics, J. of Physics.\textbf{
G} 33, 156 (2006).

\bibitem{key-2}C. Jarlskog, Phys. Rev. Lett.55, 1039 (1985).

\bibitem{key-13}N. Cabibbo, Phys.Lett \textbf{B} 72,333 (1978).

\bibitem{key-11}F. Vissani $et\, al.,$Phys.Lett. B571, 209, (2003).

\bibitem{key-12}Bipin Singh koranga, Mohan Narayan and S. Uma Sankar,
arXiv:hep-ph/0607274,(Accepted in Phys.Lett \textbf{B})

\bibitem{key-13}Bipin Singh koranga, Mohan Narayan and S. Uma Sankar,
arXiv:hep-ph/0611186..

\bibitem{key-14}Bipin Singh koranga arXiv:0707.2045.

\bibitem{key-15}Bipin Singh koranga and S. Uma Sankar, arXiv:0704.1788 
\end{thebibliography}
\end{document}